\documentclass[twocolumn,showpacs,showkeys,prc,aps]{revtex4}
\usepackage{epsfig}
\usepackage{graphicx}
\newcommand{ \be }{\begin{equation}}
\newcommand{ \ee }{\end{equation}}
\newcommand{ \bea }{\begin{eqnarray}}
\newcommand{ \eea }{\end{eqnarray}}

\begin{document}
\title{Indications for Cluster Melting from\\ Forward-Backward Charge Fluctuations at RHIC Energies}
\author{Mohamed Abdel-Aziz}
\email{abdel-aziz@th.physik.uni-frankfurt.de}
\author{Marcus Bleicher}
\affiliation{\vspace*{3mm}Institut f\"ur Theoretische Physik, J.W. Goethe Universit\"at, \\
Max von Laue Stra\ss{}e 1, 60438 Frankfurt am Main, Germany }
\date{\today}
\begin{abstract}
We study forward-backward charge fluctuations to probe the correlations among produced
particles in ultra relativistic heavy ion collisions. We develop a model that describes the
forward-backward dynamical fluctuations and apply it to interpret the recent PHOBOS data.
Within the present model, the dynamical fluctuations are related to the particle production
mechanism via cluster decay and to long range correlations between the forward and backward
rapidity hemispheres.  We argue that with a tight centrality cut, PHOBOS may see a strong
decrease of the dynamical fluctuations. Within the present model, this deterioration of  the
correlation among the produced hadrons can be interpreted as a sign for the production of  a
hot, dense and interacting medium.
\end{abstract}

%
\maketitle
%
\section{Introduction}\label{sec:introduction}
One of the main goals of the heavy ion program is to understand the nature of the hadron
production mechanism (e.g.\ parton coalescence, string fragmentation or cluster decay).
Recent RHIC data on jet quenching and elliptic flow ($v_2$) can be interpreted as an evidence
of forming a quark gluon plasma (QGP) during the collision of two heavy gold nuclei at center
of mass energy $\sqrt{s_{NN}}=200$~GeV~\cite{QM2004}. However, the detailed  mechanism how
the deconfined partonic matter transforms to hadrons is still unknown.

Using correlations and fluctuations to probe the nature of the created QCD matter has been
proposed by many authors
\cite{Baym:1995cz,Chase:1995ku,Heiselberg:1997bt,Bialas:1999tv,Voloshin:1999yf,
Stephanov:1999zu,Baym:1999up,Jeon:1999gr,Gavin:1999bk,Jeon:2000wg,Bleicher:2000ek,Heiselberg:2000ti,Bleicher:2000tr,
Dumitru:2000in,Shuryak:2000pd,Koch:2001zn,Hwa:2001xn,Abdel-Aziz:2002zn,Gavin:2001uk,Bower:2001fq,
Asakawa:2001mn,Mishustin:2005zt}.
Unfortunately, the experimental exploration of most of the suggested fluctuation signals did
not yield  positive evidence for the formation of a QGP. Nevertheless, this must not be seen
as a contradiction to the formation of a partonic state at RHIC or SPS energies, because the
proposed signatures may be weakened due to many effects (e.g.\ the phase transition is not a
strict first order transition - which is the case for RHIC energies). Other possibilities are
that rescattering and/or re-thermalization in the late QGP or in the subsequent hadronic
stage may blur the fluctuation signals. Furthermore, the fluctuation signals might be
modified and cloaked by the hadronization process itself. However, we want to point out that
recent studies of rescattering and/or diffusion in the hadronic environment, seem to indicate
that the diffusion in the hadronic stage might be small enough to allow a survival of the
initial correlations \cite{Aziz:2004qu}.

In this letter we explore the forward-backward charge fluctuations
\cite{Roland:2005ei,Steinberg:2005ec,Chai:2005fj,Wozniak:2004kp} to probe the particle
production mechanism via cluster decay. Long ago, the number $k$ of charged particles
produced per cluster decay  was estimated in pp and p$\overline{\rm p}$ interactions by the
UA5 experiment \cite{Alpgard:1983xp}. It was found that the effective multiplicity per
cluster is $k\approx 2$.  Recently, the PHOBOS experiment performed a similar analysis for
Au+Au reactions at $\sqrt{s_{NN}}=200$~GeV and reports that $k=2.7$ for peripheral collisions
and $k= 2.2$ for central collisions. Note that all measured cluster multiplicities are larger
than expected for a hadron resonance gas ($k_{\rm HG}=1.5$) \cite{Stephanov:1999zu},
indicating that the measured charge correlations can not be described by simple statistical
models based on hadronic degrees of freedom.

Our goal is to explore how the cluster structure changes if  deconfined matter is formed. The
meaning of the term cluster is rather general, for example it could be a hadronic resonance,
a partonic string or a QGP droplet. Either the survival or the destruction of the clusters
provides indirect information about the properties of the surrounding medium. For example
what is the percentage of particles produced by cluster decays, are the hadronic expectations
compatible with the data or not and if a hot and dense (partonic?) medium is formed or not.


This letter is organized as follows: In Section \ref{sec:model} we develop a model that
describes the multiplicity fluctuations $\sigma^2_C$ as measured by the PHOBOS experiment
\cite{Roland:2005ei,Steinberg:2005ec,Chai:2005fj,Wozniak:2004kp}.
This measure is sensitive to  charged particle production via cluster decay. In Section
\ref{sec:results} we apply our model to interpret the PHOBOS data and to estimate the
fraction $f$ of particles produced via cluster decay.
In Section \ref{centrality} we discuss the results of our analysis
of the measured data
\cite{Roland:2005ei,Steinberg:2005ec,Chai:2005fj,Wozniak:2004kp}
for central Au+Au collisions. Next, we address how the formation
of a dense medium affects the dynamical fluctuations because it
might influence both the cluster multiplicity and the short range
rapidity correlation length. Finally, we comment on the behavior
of the dynamical fluctuations in central collisions.
In Section \ref{sec:summary} we summarize the main results of this letter.
\section{The Model}\label{sec:model}
In this section we develop a model to describe  forward-backward
charge correlations. We define two symmetric rapidity regions at
$\pm\eta$ with equal width $\Delta\eta$. The number of charged
particles in the forward rapidity interval $\eta\pm\Delta\eta/2$
is $N_F$ while the corresponding number in the backward hemisphere
$-\eta\pm\Delta\eta/2$ is given by $N_B$. We define the
correlation variable $C$ such that
\be C=\frac{N_F-N_B}{\sqrt{N_F+N_B}}\quad.\ee
The event-by-event fluctuations (variance) $\sigma^2_C$ of $C$ are decomposed into two parts,
weighted by the respective fractions $f$ and $(1-f)$ of particles from the different sources:
\be\label{eq:definition}
 \sigma^2_C=f \sigma^2_{\rm SR}+(1-f)\sigma^2_{\rm LR}\quad.
 \ee
The first term in Eq. (\ref{eq:definition}) $\sigma^2_{\rm SR}$ denotes the short range
correlation due to cluster decays. This term is a direct measure of the particle production
via cluster decay. As discussed above, the specific nature of the clusters is rather general.

$\sigma^2_{\rm SR}$  depends on  the observability of the decay
daughters in the rapidity width $\Delta\eta$, so that when
$\Delta\eta$ is larger than the size of the cluster one is able to
observe all the decay products of this cluster. However, if
$\Delta\eta$  is small one may ``loose'' the decay products of
such a cluster. To model this rapidity window effect we propose
the following form
\be \label{window}\sigma^2_{\rm SR}=k\left[1-\exp({-\Delta\eta/\lambda_{\rm
short}})\right]\quad,\ee
where $\lambda_{\rm short}$ is the rapidity correlation length between particles produced as
a result of the cluster decay and $k$ is the aforementioned multiplicity per cluster.

The second term $\sigma^2_{\rm LR}$ in equation (\ref{eq:definition}) includes the
fluctuations due to long range correlations between particles in both hemispheres and the
background particles that are statistically independent.
To describe the long range correlation, it is convenient to introduce a two body distribution
function $\rho_2(\eta_1,\eta_2)$. Such that
\be \rho_2(\eta_1,\eta_2)=A \exp\left[{-\frac{(\eta_1-\eta_2)^2}{2\lambda_{\rm
long}^2}}\right]\quad,\ee
where $A$ is the amplitude of the correlation function, giving the strength of the two body
distribution when $\eta_1=\eta_2$. Generally, $A$ is a product of the true amplitude and a
function that might depend on the motion of the center of mass rapidity of the pair
$\eta_c=(\eta_1+\eta_2)/2$. For the present study, we  assume that $A$ is independent of
$\eta$ and $\Delta \eta$.  Under these assumptions, the long range contribution to the total
fluctuations can be calculated from
\be\label{eq:main1} \sigma^2_{_{\rm
LR}}=1-A\int\limits^{-\eta+\Delta\eta/2}_{-\eta-\Delta\eta/2}{\rm d}\eta_1
\int\limits^{\eta+\Delta\eta/2}_{\eta-\Delta\eta/2}{\rm
d}\eta_2~\exp\left[{-\frac{(\eta_1-\eta_2)^2}{2\lambda_{\rm long}^2} }\right].\ee
Next,  we decompose the total multiplicity $N$ in each rapidity window $\Delta\eta$  in two
parts such that $N=N_{\rm SR}+N_{\rm LR}$ where $N_{\rm SR}$ is the charged particle
abundance produced from the decay of all clusters inside $\Delta\eta$ and $N_{\rm LR}$ is the
number of particles migrated from outside into the rapidity window under investigation.

In each rapidity bin $\Delta\eta$, the fraction of particles produced by cluster decay $f$ is
given  by $f=N_{\rm SR}/N$. With the weighting of  $\sigma^2_{\rm SR}$ and $\sigma^2_{\rm
LR}$ by $f$ and $(1-f)$, and the abbreviation  $\xi=(1-f)A$ one rewrites Equation
(\ref{eq:definition}) using Eqs. (\ref{window}) and (\ref{eq:main1}) as
\be\label{eq:main} \sigma^2_C = 1+f\left[k\left(1- {\rm e}^{ -\frac{\Delta\eta}{\lambda_{\rm
short}}}\right)-1\right] - \xi F(\lambda_{\rm long},\Delta\eta,\eta), \ee
where $F(\lambda_{\rm long},\Delta\eta,\eta)$ is the result of the double integral in
Equation (\ref{eq:main1}).
Equation (\ref{eq:main}) includes different physical situations, e.g. in case of vanishing
short range correlation $f=0$,  and $\sigma^2_C$ is determined solely by long range
correlations. In case of $f=1$, only short range correlations (i.e. cluster decays)
contribute to the dynamical fluctuations. Therefore, if  measurements are done such that the
two bins centered around $\pm \eta$  are far from each other, the contribution from long
range correlations becomes negligible. Only in this case, an increase of the bin size
$\Delta\eta$ to $\Delta\eta\gg\lambda_{\rm short}$ leads to the commonly used approximation
$\sigma^2_C\approx k$ from which the true cluster multiplicity can be deduced.  Another
limiting value obeyed by Eq. (\ref{eq:main}) is that for a very large $\eta$ the dynamical
fluctuations $\sigma^2_C$ do approach the poisson value $\sigma^2_C=1$. This is due to the
vanishing of the long range correlations and the low multiplicity which drives the fraction
of correlated particles to be zero.

Before we analyze the PHOBOS data let us shortly summarize the assumptions used in the above
derivations: (I) There is only one type of clusters. (II) All clusters  decay to the same
number of particles $k$, i.e. fluctuations of $k$ itself are not included in the present
model.

\section{Results}\label{sec:results}
\begin{figure}
\centerline{\epsfig{file=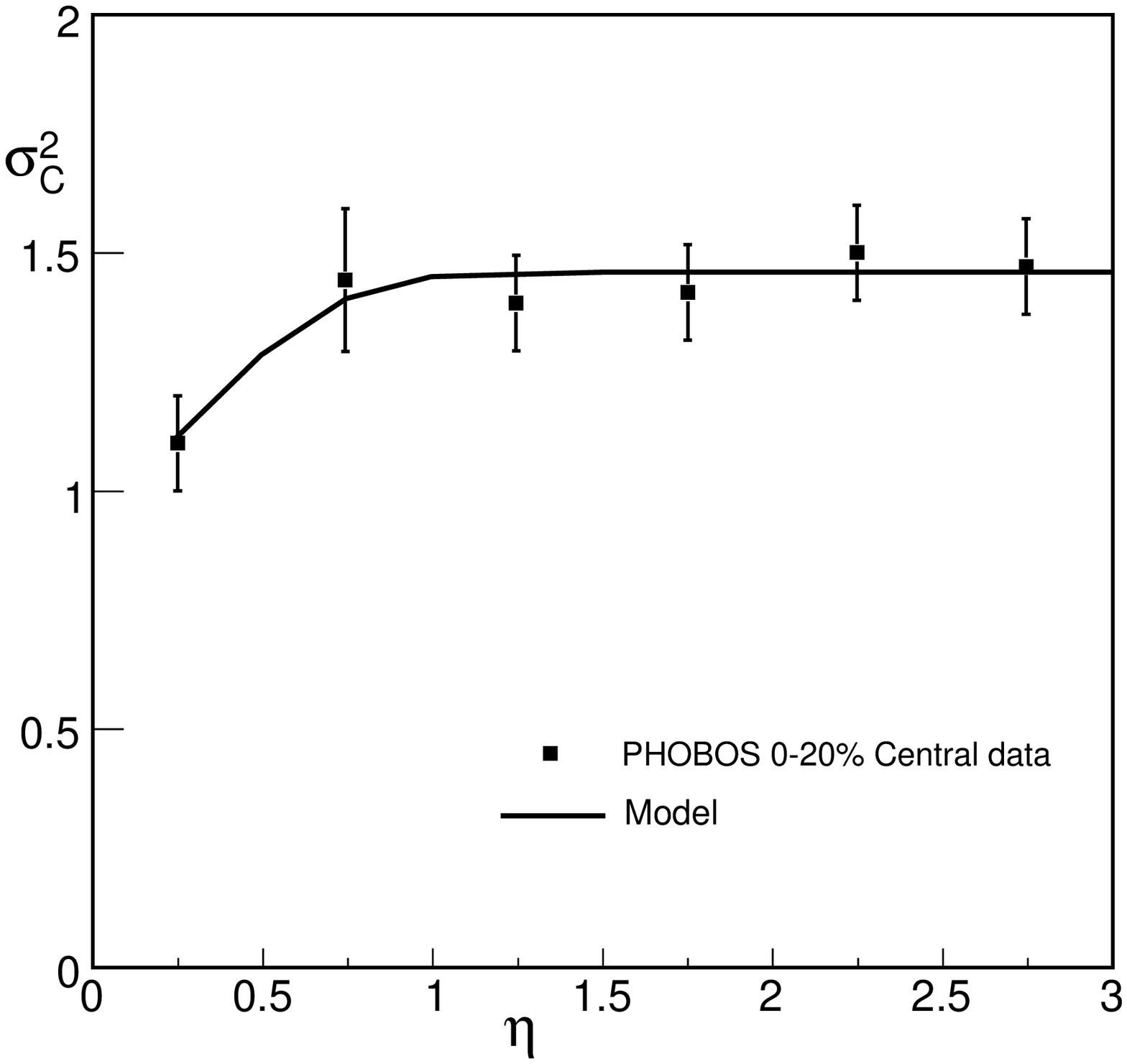,width=8cm}} \vskip -0.4in\caption{$\sigma^2_C$
versus rapidity $\eta$ for $0\%-20\%$ central Au+Au at $\Delta\eta=2$. Black squares are
PHOBOS data \cite{Roland:2005ei,Steinberg:2005ec,Chai:2005fj,Wozniak:2004kp}. The line is the
model calculation  using equation (\ref{eq:main}).} \label{fig:central1}\end{figure}

In this section we study the dynamical fluctuations $\sigma^2_C$ for Au+Au collisions at
$\sqrt{s_{NN}}=200$~GeV. We extract the values of $f$ and $\xi$ from PHOBOS data reported in
\cite{Roland:2005ei,Steinberg:2005ec,Chai:2005fj,Wozniak:2004kp}. Table \ref{table:1}
summarizes our findings on the parameters of the present model. With these parameters, we
predict $\sigma^2_C$ as a function of $\eta$ and $\Delta\eta$. Figure \ref{fig:central1}
depicts both PHOBOS data for $0\%-20\%$ central Au+Au collisions (symbols) and our model
results (line) for the dynamical fluctuations~$\sigma^2_C$ as a function of $\eta$ while the
width of the observation window $\Delta\eta$ is kept as $\Delta\eta=0.5$.

\begin{figure}
\centerline{\epsfig{file=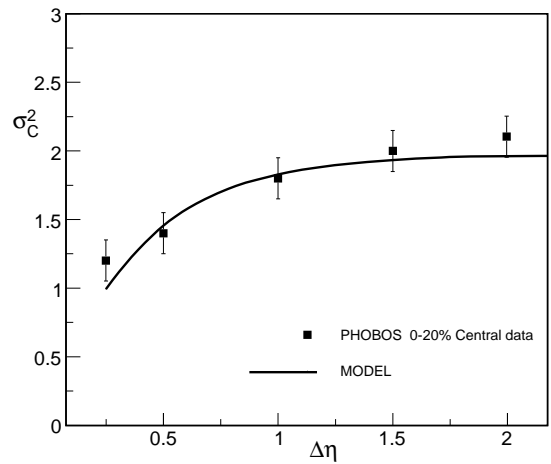,width=8cm}} \vskip -0.4in\caption{$\sigma^2_C$ as a
function of $\Delta\eta$ for $0\%-20\%$ central Au+Au at $\eta=2$. Black squares are PHOBOS
data \cite{Roland:2005ei,Steinberg:2005ec,Chai:2005fj,Wozniak:2004kp}. The line is the model
calculation using equation (\ref{eq:main}).} \label{fig:central2}\end{figure}

In Fig. \ref{fig:central2} we show $\sigma^2_C$ versus $\Delta\eta$, while the center of the
observation window is fixed at $\eta=2$. We emphasize that the same parameter set (see Table
\ref{table:1}) is used for both data sets in Figs. \ref{fig:central1} and \ref{fig:central2}.

\begin{figure}
\centerline{\epsfig{file=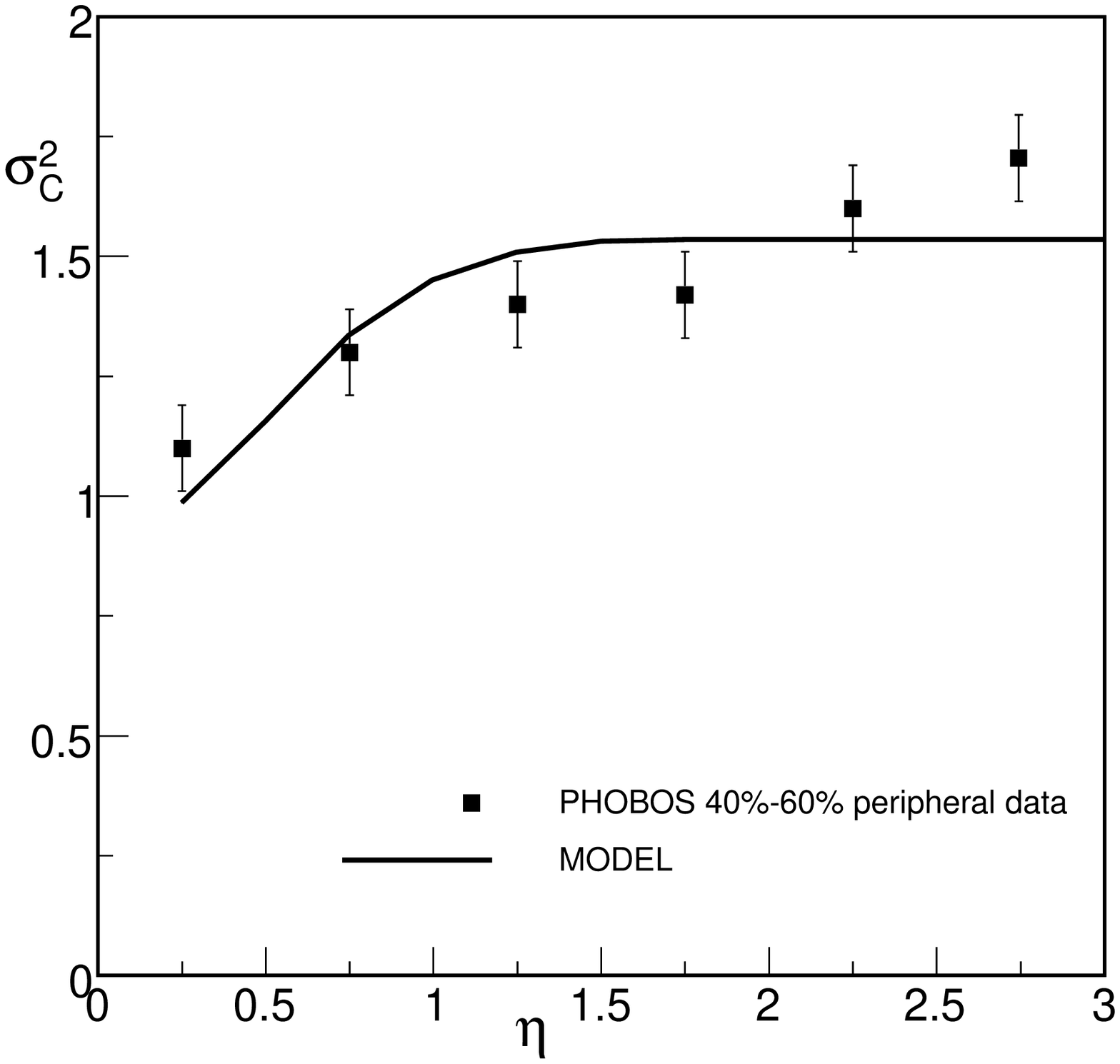,width=8cm}} \vskip -0.4in\caption{$\sigma^2_C$
versus rapidity $\eta$ for $40\%-60\%$ peripheral Au+Au at $\Delta\eta=0.5$. Black squares
are PHOBOS data \cite{Roland:2005ei,Steinberg:2005ec,Chai:2005fj,Wozniak:2004kp}. The line is
the model calculation using equation (\ref{eq:main}).} \label{fig:figure5}\end{figure}

\begin{figure}
\centerline{\epsfig{file=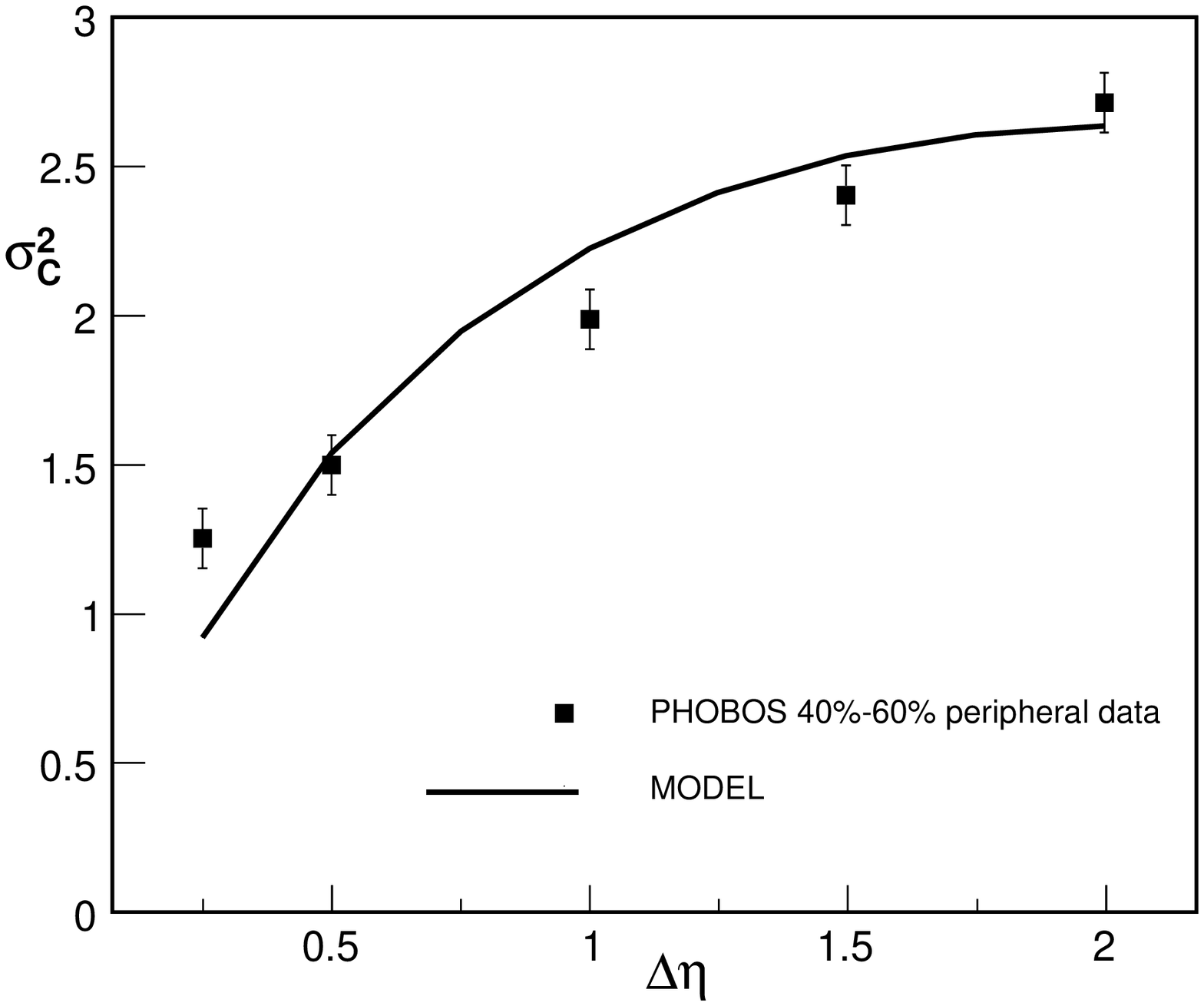,width=8cm}} \vskip -0.4in\caption{$\sigma^2_C$
versus $\Delta\eta$ for $40\%-60\%$ peripheral Au+Au at $\eta=2$. Black squares are PHOBOS
data \cite{Roland:2005ei,Steinberg:2005ec,Chai:2005fj,Wozniak:2004kp}. The line is the model
calculation using equation (\ref{eq:main}).} \label{fig:figure6}\end{figure}

Before we discuss the interpretation of our model results, let us turn to semi-peripheral
collisions. Figs.~\ref{fig:figure5} and \ref{fig:figure6} depict the PHOBOS results on
$40\%-60\%$ peripheral collisions. As for central collisions, the model allows to describe
both data samples with the same parameter set simultaneously (cf. Table \ref{table:1}). The
dynamical fluctuations $\sigma^2_C$ measured at $\eta=2$ for $40\%-60\%$ peripheral Au+Au
collisions and a very wide rapidity windows $\Delta\eta=2$ yield an effective cluster
multiplicity of $k\approx 2.7$ (see Fig. \ref{fig:figure6}).

One observes that the short range correlation length $\lambda_{\rm
short}$ is 0.6  for peripheral collisions and decreases towards
$\lambda_{\rm short}=0.4$ in central collisions.  This decrease of
the short range correlation length is consistent with Ref.
\cite{Jeon:2005yi,Jeon:2005kj,Shi:2005rc} and was speculated to be
a signal for the formation of a quark gluon plasma. The fraction
of correlated particles decreases from almost 100\% in peripheral
collisions to 88\% for central reactions.

\begin{center}
\begin{table}
\begin{tabular}{|c|c|c|c|c|}
\hline
data & $f$ & $\xi$ & $\lambda_{\rm short}$ & $\lambda_{\rm long}$ \\
\hline
 Au+Au $0\%-20\%$ & 0.88 & 1.8 & 0.4 & 0.7\\
\hline
 Au+Au $40\%-60\%$&0.99 & 2.5 & 0.6 &0.9\\
 \hline
\end{tabular}
\caption{Parameters of our model as estimated by analyzing $0\%-20\%$ central and $40\%-60\%$
peripheral Au+Au PHOBOS data at $\sqrt{s_{\rm NN}}=200$~GeV using equation (\ref{eq:main}).}
\label{table:1}
\end{table}
\end{center}
\vskip -0.4in

%
\section{centrality dependance of $\sigma^2_C$}\label{centrality}
\begin{figure}
\centerline{\epsfig{file=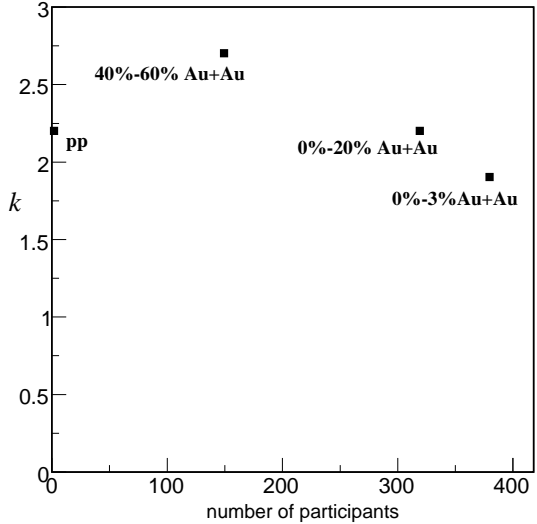,width=8cm}} \vskip
-0.15in\caption{Cluster multiplicity $k$ versus the number of
participants. The last point represents 0\%-3\% Au+Au is computed
by extrapolating our model parameters in Table \ref{table:1}.}
\label{fig:centrality}\end{figure}
In Fig. \ref{fig:centrality} we show the experimental results from the UA5 and the PHOBOS
experiment on the fluctuations of the forward-backward correlation $\sigma^2_C$ for
p$\overline{\rm p}$  (Au+Au) at different centralities.

From Fig. \ref{fig:centrality}, we see that $\sigma^2_C$ starts at
2.2 for p$\overline{\rm p}$ and then increases to 2.7 for
$40\%-60\%$ peripheral and then goes down to 2.2 again for
$0\%-20\%$ central Au+Au. This behavior has important consequences
that will help to understand the particles productions mechanism
in nucleus-nucleus collisions compared to nucleon-nucleon
collisions as we will discuss next.

The increase in the dynamic fluctuations for $40\%-60\%$ peripheral collisions compared to
p$\overline{\rm p}$ at the same energy might be attributed to the difference in the dynamics
of the collisions between p$\overline{\rm p}$ and nucleus-nucleus collisions. First let us
consider the string picture for particles production in p$\overline{\rm p}$ collisions.
During such collisions a few strings between the quarks in both protons form and are
subsequently fragmented into hadrons. If the same picture holds for nucleus-nucleus
collisions, then $\sigma^2_C$ in both p$\overline{\rm p}$ and Au+Au will be similar and
$\sigma^2_C$ has a flat centrality dependence. However, experimental data indicates that the
cluster multiplicity $k$ is increased in $40\%-60\%$ peripheral Au+Au collisions compared to
p$\overline{\rm p}$ interactions. An explanation of this enhancement might be found in the
context of string fusion models  \cite{Braun:1997ch}. Here, the fusion process might increase
the probability of high mass clusters compared to p$\overline{\rm p}$ collisions, resulting
in a higher multiplicity per cluster.

If the formation of heavier clusters is the explanation for the increase of $k$ in Au+Au
compared to p$\overline{\rm p}$, then one expects an even further increase of $k$ towards
central Au+Au collisions. This enhancement will be observable if interactions with the
surrounding medium are negligible compared to the increase in $k$ due to string fusion. Fig.
5 shows that this is not the case, because the cluster multiplicity in $0\%-20\%$ central
Au+Au collisions decreases  to $k=2.2$ . In addition, the number of correlated particles $f$
also decreases towards more central Au+Au reactions (cf. Table \ref{table:1}). This indicates
that in central collisions the interactions with the hot and dense QCD matter results in
smaller clusters (reducing the value of $k$) as compared to peripheral collisions. With even
tighter centrality cuts one might be able to see a complete melting of the clusters leading
to independent uncorrelated particle emission patterns (i.e.\ $\sigma^2_C=1$ and $k=1$).

To test this idea, we linearly extrapolate those parameters in
Table \ref{table:1} to $0\%-3\%$ Au+Au collisions. The last point
in Fig. \ref{fig:centrality} corresponds to this centrality cuts.
We find that $k\approx 1.9, \lambda_{\rm short}\approx 0.3$. This
means that clusters formed at the most central collisions are
partially melted compared to mid-peripheral collisions. Also the
short range correlation length $\lambda_{\rm short}$ becomes
smaller. It is worth to mention that the decrease of the rapidity
correlation length seems to be a general feature of
nucleus-nucleus collisions at high energies since the width of the
balance function expected to decrease with centrality as a result
of late hadronization as mentioned in Ref. \cite{Bass:2000az}.
This decrease is observed in the data reported in
\cite{Alt:2004gx,Adams:2003kg}.

To distinguish between hadronic and QGP effects, $\sigma^2_C$ should be measured at different
energies and ion sizes and also with different transverse momentum  cuts to include/exclude
the effects of mini-jets, since jets hadronize into a shower of hadrons which might mimic a
clustering effect (within the jet cone). Thus,  limiting the measurement of $\sigma^2_C$ to
the soft region (i.e. $p_t<2$~GeV) allows to minimize jet artifacts.

In this case conclusive statement can be given by comparing peripheral data to the central
one to extract $f$ and $k$ for different initial conditions and to study how both $f$ and $k$
change as a function of the initial partonic volume, life time and energy density.
\section{Summary and Conclusion}\label{sec:summary}
In this letter we present a model to analyze forward-backward dynamical fluctuations as
recently measured by the PHOBOS collaboration in Au+Au reactions at $\sqrt {s_{\rm
NN}}=200$~GeV. We find that the effective cluster multiplicity in central collisions is lower
than for peripheral collisions. Also the short range rapidity correlation length decreases
towards more central interactions.

This might indicate that the clusters are ``melted'' in central collisions, resulting in a
smaller fraction of correlated particles (decrease of $f$). One can test the degree of
melting by measuring $k$ as a function of centrality, and especially for  very central Au+Au
collisions. If one observes that $k\rightarrow 1$, all clusters are melted and the final
particles are independently emitted.
\section*{Acknowledgements}
We thank S. Haussler and  A. Kostyuk for fruitful discussions and valuable comments. This
work is supported by BMBF, GSI and DAAD.
%



\begin{thebibliography}{99}
\bibitem{QM2004}QUARK MATTER 2004 - 17th International
Conference On Ultra-Relativistic Nucleus-Nucleus Collisions
(OAKLAND, CALIFORNIA, USA, 11-17 JANUARY 2004);
Quark Matter 2005,The 18th International Conference in
Nulcues-Nucleus Collisions, August 4-9, Budapest, Hungary.

\bibitem{Baym:1995cz}
  G.~Baym, B.~Blattel, L.~L.~Frankfurt, H.~Heiselberg and M.~Strikman,
  Phys.\ Rev.\ C {\bf 52}, 1604 (1995)
  [arXiv:nucl-th/9502038].

\bibitem{Chase:1995ku}
  K.~C.~Chase and A.~Z.~Mekjian,
  Phys.\ Lett.\ B {\bf 379}, 50 (1996)
  [arXiv:nucl-th/9510005].

\bibitem{Heiselberg:1997bt}
  H.~Heiselberg and A.~P.~Vischer,
  Phys.\ Lett.\ B {\bf 421}, 18 (1998)
  [arXiv:nucl-th/9703030].


\bibitem{Bialas:1999tv}
  A.~Bialas and V.~Koch,
  Phys.\ Lett.\ B {\bf 456}, 1 (1999)
  [arXiv:nucl-th/9902063].

\bibitem{Voloshin:1999yf}
  S.~A.~Voloshin, V.~Koch and H.~G.~Ritter,
  Phys.\ Rev.\ C {\bf 60}, 024901 (1999)
  [arXiv:nucl-th/9903060].

\bibitem{Stephanov:1999zu}
  M.~A.~Stephanov, K.~Rajagopal and E.~V.~Shuryak,
  Phys.\ Rev.\ D {\bf 60}, 114028 (1999)
  [arXiv:hep-ph/9903292].


\bibitem{Baym:1999up}
  G.~Baym and H.~Heiselberg,
  Phys.\ Lett.\ B {\bf 469}, 7 (1999)
  [arXiv:nucl-th/9905022].

\bibitem{Jeon:1999gr}
  S.~Jeon and V.~Koch,
  Phys.\ Rev.\ Lett.\  {\bf 83}, 5435 (1999)
  [arXiv:nucl-th/9906074].


\bibitem{Gavin:1999bk}
  S.~Gavin,
  arXiv:nucl-th/9908070.

\bibitem{Jeon:2000wg}
  S.~Jeon and V.~Koch,
  Phys.\ Rev.\ Lett.\  {\bf 85}, 2076 (2000)
  [arXiv:hep-ph/0003168].

\bibitem{Bleicher:2000ek}
  M.~Bleicher, S.~Jeon and V.~Koch,
  Phys.\ Rev.\ C {\bf 62}, 061902 (2000)
  [arXiv:hep-ph/0006201].

\bibitem{Heiselberg:2000ti}
  H.~Heiselberg and A.~D.~Jackson,
  Phys.\ Rev.\ C {\bf 63}, 064904 (2001)
  [arXiv:nucl-th/0006021].

\bibitem{Bleicher:2000tr}
  M.~Bleicher, J.~Randrup, R.~Snellings and X.~N.~Wang,
  Phys.\ Rev.\ C {\bf 62}, 041901 (2000)
  [arXiv:nucl-th/0006047].

\bibitem{Dumitru:2000in}
  A.~Dumitru and R.~D.~Pisarski,
  Phys.\ Lett.\ B {\bf 504}, 282 (2001)
  [arXiv:hep-ph/0010083].

\bibitem{Shuryak:2000pd}
  E.~V.~Shuryak and M.~A.~Stephanov,
  Phys.\ Rev.\ C {\bf 63}, 064903 (2001)
  [arXiv:hep-ph/0010100].
\bibitem{Koch:2001zn}
  V.~Koch, M.~Bleicher and S.~Jeon,
  Nucl.\ Phys.\ A {\bf 698}, 261 (2002)
  [Nucl.\ Phys.\ A {\bf 702}, 291 (2002)]
  [arXiv:nucl-th/0103084].
%
\bibitem{Hwa:2001xn}
  R.~C.~Hwa and C.~B.~Yang,
  Phys.\ Lett.\ B {\bf 534}, 69 (2002)
  [arXiv:hep-ph/0104216].
\bibitem{Abdel-Aziz:2002zn}
  M.~Abdel-Aziz and S.~Gavin,
  Nucl.\ Phys.\ A {\bf 715}, 657 (2003)
  [J.\ Phys.\ G {\bf 30}, S271 (2004)]
  [arXiv:nucl-th/0209019].

\bibitem{Gavin:2001uk}
  S.~Gavin and J.~I.~Kapusta,
  Phys.\ Rev.\ C {\bf 65}, 054910 (2002)
  [arXiv:nucl-th/0112083].

\bibitem{Bower:2001fq}
  D.~Bower and S.~Gavin,
  Phys.\ Rev.\ C {\bf 64}, 051902 (2001)
  [arXiv:nucl-th/0106010].
%
%
\bibitem{Asakawa:2001mn}
  M.~Asakawa, U.~W.~Heinz and B.~Muller,
  Nucl.\ Phys.\ A {\bf 698}, 519 (2002)
  [arXiv:nucl-th/0106046].
%
\bibitem{Mishustin:2005zt}
  I.~N.~Mishustin,
  arXiv:hep-ph/0512366.

\bibitem{Aziz:2004qu}
  M.~A.~Aziz and S.~Gavin,
  Phys.\ Rev.\ C {\bf 70}, 034905 (2004)
  [arXiv:nucl-th/0404058].

\bibitem{Roland:2005ei}
  G.~Roland \textit{et al.}  [PHOBOS Collaboration],
  arXiv:nucl-ex/0510042.
%
\bibitem{Steinberg:2005ec}
  P.~Steinberg \textit{et al.}  [PHOBOS Collaboration],
  arXiv:nucl-ex/0510036.
%
\bibitem{Chai:2005fj}
  Z.~w.~Chai \textit{et al.}  [PHOBOS Collaboration],
  arXiv:nucl-ex/0509027.
%
%
\bibitem{Wozniak:2004kp}
  K.~Wozniak \textit{et al.}  [PHOBOS Collaboration],
  J.\ Phys.\ G {\bf 30}, S1377 (2004).


\bibitem{Alpgard:1983xp}
  K.~Alpgard \textit{et al.}  [UA5 Collaboration],
  Phys.\ Lett.\ B {\bf 123}, 361 (1983);
  R.~E.~Ansorge {\it et al.}  [UA5 Collaboration],
  Z.\ Phys.\ C {\bf 37}, 191 (1988).

\bibitem{Jeon:2005yi}
  S.~Jeon, L.~Shi and M.~Bleicher,
  J.\ Phys.\ Conf.\ Ser.\  {\bf 27}, 194 (2005)
  [arXiv:nucl-th/0511066].
\bibitem{Shi:2005rc}
  L.~j.~Shi and S.~Jeon,
  Phys.\ Rev.\ C {\bf 72}, 034904 (2005)
  [arXiv:hep-ph/0503085].
%
\bibitem{Jeon:2005kj}
  S.~Jeon, L.~Shi and M.~Bleicher,
  Phys.\ Rev.\ C {\bf 73}, 014905 (2006)
  [arXiv:nucl-th/0506025].

\bibitem{Braun:1997ch}
  M.~A.~Braun, C.~Pajares and J.~Ranft,
  Int.\ J.\ Mod.\ Phys.\ A {\bf 14}, 2689 (1999)
  [arXiv:hep-ph/9707363].

\bibitem{Bass:2000az}
  S.~A.~Bass, P.~Danielewicz and S.~Pratt,
  Phys.\ Rev.\ Lett.\  {\bf 85}, 2689 (2000)
  [arXiv:nucl-th/0005044].

\bibitem{Alt:2004gx}
  C.~Alt {\it et al.}  [NA49 Collaboration],
  Phys.\ Rev.\ C {\bf 71}, 034903 (2005)
  [arXiv:hep-ex/0409031].
  %
\bibitem{Adams:2003kg}
  J.~Adams {\it et al.}  [STAR Collaboration],
  Phys.\ Rev.\ Lett.\  {\bf 90}, 172301 (2003)
  [arXiv:nucl-ex/0301014].


\end{thebibliography}
\end{document}